\documentclass[12pt]{article}
\title{Analytic solution for Brownian motion in three dimensional hyperbolic space}
\author{ 
Naomichi \textsc{Suzuki}$^{1,}$\footnote{E-mail:suzuki@matsu.ac.jp} 
and Minoru \textsc{Biyajima}$^{2,}$\footnote{E-mail:mbiyajima@azusa.shinshu-u.ac.jp}  \\ 
$^1$Department of Comprehensive Management, Matsumoto University, \\
 Matsumoto 390-1295, Japan\\
$^2$Department of Physics, Shinshu University, Matsumoto, 390-8621, Japan
}

\begin{document}
\date{}
\maketitle

 \begin{abstract}
Brownian motion in the three dimensional Lobachevsky space or hyperbolic space is considered in the paper written by
F.I.Karpelevich, V.N.Tutubalin and M.G.Shur.   A solution for radial symmetric diffusion equation 
in the three dimensional hyperbolic space is given in that paper. However, derivation of it is not explicitly shown.
Therefore, the diffusion equation is solved analytically with an initial condition corresponding to non-zero space variable 
at the initial stage.
 \end{abstract}

\section{Introduction}

A radial part of Brownian motion in the three dimensional Lobachevsky space or hyperbolic space is written by the diffusion equation, 
\begin{eqnarray*}
  \frac{\partial f}{\partial t}= \frac{D}{\sinh^2\!\rho}\, 
      \frac{\partial}{\partial \rho}\left( 
        \sinh^2\!\rho\, \frac{\partial f}{\partial \rho} 
      \right).          \label{eq.int1}
\end{eqnarray*}
A solution of the diffusion equation is  given in Ref.~\cite{karp59} as
\begin{eqnarray}
  f(\rho,t) = \frac{1}{\sqrt{(4\pi Dt)^3}}{\rm e}^{-Dt} \frac{\rho}{\sinh\rho}
                \exp[-\frac{\rho^2}{4Dt}],    \label{eq.int3}
\end{eqnarray}
where $D$ denotes a diffusion constant.

We have applied Eq.(\ref{eq.int3}) to the analyses of large transverse momentum distributions of secondary particles observed in high energy nucleus-nucleus ($AA$) collisions~\cite{suzu04}. 
We consider $\rho$ in Eq.(\ref{eq.int3}) as radial rapidity, which is written with energy $E$, momentum ${\bf p}$ and mass $m$ of observed particle by
\begin{eqnarray*}
  \rho=\ln\frac{E+|{\bf p}|}{m}.
\end{eqnarray*}
Transverse momentum $ p_T$ is defined by $ p_T=|{\bf p}|\sin \theta$, where  $\theta$ is the polar angle of the observed particle measured from the direction of incident colliding nuclei.\footnote{
In high energy nucleus-nucleus collisions, colliding energy $\sqrt{s}$ grows up to  $\sqrt{s}=200$ GeV, and secondary particles with transverse momentum $p_T$ more than 10 GeV/c are observed. The observed $p_T$ distributions have long tail compared with exponential distribution in $p_T$. As is well known, the fundamental solution of stochastic process is Gaussian if variables in the euclidian space are used.  Therefore, as long as we consider stochastic equations in the transverse momentum space, it is very hard to describe observed $p_T$ distributions. This fact suggests that a relativistic approach to the stochastic process would be needed.
}

However, derivation of solution (\ref{eq.int3}) is not explicitly shown in Ref.~\cite{karp59}.  In addition, up to the present, we cannot find the article where the analytical proof is made.\footnote{In Ref.~\cite{wata75}, the solution is obtained from a Markovian property of Brownian motion. }  
Therefore, we would like to solve the diffusion equation analytically.

\section{Diffusion equation in three dimensional hyperbolic space}

The radial symmetric diffusion equation in the three dimensional hyperbolic space $H_3$ is given by,
\begin{eqnarray}
  \frac{\partial f}{\partial t}= \frac{D}{\sinh^2\!\rho}\, 
      \frac{\partial}{\partial \rho}\left( 
        \sinh^2\!\rho\, \frac{\partial f}{\partial \rho} 
      \right).          \label{eq.app1}
\end{eqnarray}
The initial condition for Eq.(\ref{eq.app1}) is taken as,
\begin{eqnarray}
    f(\rho,\rho_0,t=0)= \frac{\delta(\rho-\rho_0)}{4\pi\sinh^2\!\rho}.   \label{eq.app2}
\end{eqnarray}
At first,  $f(\rho,\rho_0,t)$ is put as
\begin{eqnarray}
    f(\rho,\rho_0,t)= \frac{1}{\sinh\rho}{\rm e}^{-Dt} g(\rho,t).     \label{eq.app3}
\end{eqnarray}
From Eqs.(\ref{eq.app1}) and (\ref{eq.app3}), equation for $g(\rho,t)$ is given by 
\begin{eqnarray}
   \frac{\partial g}{\partial t}
      -D\frac{\partial^2 g}{\partial \rho^2}=0,          \label{eq.app4}
\end{eqnarray}
which is the diffusion equation in one dimension.  The initial condition for  Eq.(\ref{eq.app4}) is given from Eqs.(\ref{eq.app2}) and (\ref{eq.app3}) as,
\begin{eqnarray}
    g(\rho,t=0)= \frac{\delta(\rho-\rho_0)}{4\pi\sinh\rho}.    \label{eq.app5}
\end{eqnarray}
The Laplace transform,  
\begin{eqnarray}
  G(\rho,s) = \mathcal{L}[g(\rho,t),s] = \int_0^\infty g(\rho,t) {\rm e}^{-st}dt,
         \label{eq.app6}
\end{eqnarray}
is applied to Eq.(\ref{eq.app4});
\begin{eqnarray}
   \frac{d^2 G(\rho,s)}{d \rho^2} - \kappa^2G(\rho,s) 
   &=& -\frac{1}{D} g(\rho,t=0), \nonumber  \\
   \kappa &=& \sqrt{{s}/{D}}.
      \label{eq.app7}
\end{eqnarray}
In order to solve Eq.(\ref{eq.app7}), we put
\begin{eqnarray}
  G(\rho,s)=C(\rho){\rm e}^{-\kappa \rho}.  \label{eq.app8}
\end{eqnarray}
From Eqs.(\ref{eq.app7}) and (\ref{eq.app8}), we obtain
\begin{eqnarray}
  \frac{d^2 C(\rho)}{d\rho^2} - 2\kappa \frac{d C(\rho)}{d \rho} 
  = \frac{1}{D} g(\rho,t=0){\rm e}^{\kappa \rho}.    \label{eq.app9}
\end{eqnarray} 
Equation (\ref{eq.app9}) is rewritten as,
\begin{eqnarray*}
   \frac{d }{d \rho} \left(  \frac{d C(\rho)}{d \rho}{\rm e}^{-2 \kappa\rho} \right)
   = -\frac{1}{D} g(\rho,t=0){\rm e}^{-\kappa\rho}. 
      \label{eq.app12}
\end{eqnarray*}
After integrating over $\rho$, we obtain,
\begin{eqnarray*}
     \frac{d C(\rho)}{d \rho} = B(0){\rm e}^{2 \kappa\rho} 
       -\frac{1}{D}{\rm e}^{2 \kappa\rho}\int_0^\rho
        g(\xi,t=0){\rm e}^{-\kappa\xi} d\xi, \label{eq.app13}
\end{eqnarray*}
where $B(0)=d C(0)/d \rho$. Then, $C(\rho)$ is given in the following form,
\begin{eqnarray}
   C(\rho) &=& C(0) +  \frac{B(0)}{2\kappa}
                      \left({\rm e}^{2 \kappa\rho}-1\right) 
   -\frac{1}{D}\int_0^\rho {\rm e}^{2 \kappa\rho^\prime}d\rho^\prime
        \int_0^{\rho^\prime} g(\xi,t=0){\rm e}^{-\kappa\xi} d\xi,  \nonumber \\
   &=& C(0) + \frac{B(0)}{2\kappa}\left({\rm e}^{2 \kappa\rho}-1 \right) 
   -\frac{1}{2\kappa D}\int_0^\rho g(\xi,t=0)
    \left({\rm e}^{2\kappa\rho-\kappa\xi}-{\rm e}^{\kappa\xi}\right) d\xi.
             \label{eq.app14}
\end{eqnarray}
If $\rho>\rho_0$, after substituting the initial condition (\ref{eq.app5}) to Eq.(\ref{eq.app14}), we have an expression for $C(\rho)$;  
\begin{eqnarray*}
   C(\rho) = C(0) + \frac{{\rm e}^{2 \kappa\rho}}{2\kappa}    
       \left( B(0)-\frac{{\rm e}^{-\kappa\rho_0}}{4\pi D \sinh \rho_0} \right)
      + \frac{1}{2\kappa} 
     \left( \frac{{\rm e}^{\kappa\rho_0}}{4\pi D \sinh \rho_0}-B(0) \right).
             \label{eq.app15}
\end{eqnarray*}
Then, $G(\rho,s)$, the Laplace transform of $g(\rho,t)$ is written as
\begin{eqnarray*}
   G(\rho,s)&=& C(\rho){\rm e}^{-\kappa\rho}  \nonumber \\
   &=& C(0){\rm e}^{-\kappa\rho}
         + \frac{{\rm e}^{\kappa\rho}}{2\kappa}
        \left( B(0)-\frac{{\rm e}^{-\kappa\rho_0}}{4\pi D \sinh \rho_0} \right)
      + \frac{{\rm e}^{-\kappa\rho}}{2\kappa} 
     \left( \frac{{\rm e}^{\kappa\rho_0}}{4\pi D \sinh \rho_0}-B(0) \right).
\end{eqnarray*}
In order to get inverse Laplace transform of $G(\rho,s)$,  equation,
\begin{eqnarray*}
   B(0)=\frac{{\rm e}^{-\kappa\rho_0}}{4\pi D \sinh \rho_0},  
\end{eqnarray*}
should be satisfied.
Then, we have
\begin{eqnarray}
   G(\rho,s)=C(0){\rm e}^{-\kappa\rho}  + \frac{1}{4\pi D\kappa} 
          \frac{\sinh \kappa\rho_0}{ \sinh \rho_0}{\rm e}^{-\kappa\rho}.
   \label{eq.app16}
\end{eqnarray}
Substituting $\kappa=\sqrt{s/D}$ to Eq.(\ref{eq.app16}), we obtain,

\begin{eqnarray*}
   G(\rho,s)=C(0){\rm e}^{-{\rho}\sqrt{s}/{\sqrt{D}}} 
        + \frac{1}{4\pi D}\frac{1}{ \sinh \rho_0}
          \sum_{m=0}^\infty \left(\frac{s}{D}\right)^m 
          \frac{\rho_0^{2m+1}}{(2m+1)!}{\rm e}^{-{\rho}\sqrt{s}/{\sqrt{D}}} .
   \label{eq.app17}
\end{eqnarray*}

The inverse Laplace transform of $G(\rho,s)$ is defined by,
\begin{eqnarray*}
  g(\rho,t)=\mathcal{L}^{-1}[G(\rho,s);t]=\frac{1}{2\pi i}
      \int_{c-i\infty}^{c+i\infty}G(\rho,s)\exp[st] ds,
\end{eqnarray*}
where $c$ is a constant.  
The following formulae for the inverse Laplace transform\cite{abra72},
\begin{eqnarray*}
  \mathcal{L}^{-1} [ \exp[-{\rho}\sqrt{s}/{\sqrt{D}}];t]
    &=& \frac{\rho}{\sqrt{4\pi Dt^3}} \exp[-\frac{\rho^2}{4Dt}],    \nonumber \\
  \mathcal{L}^{-1}[s^m\exp[-{\rho}\sqrt{s}/{\sqrt{D}}];t] 
    &=& \frac{1}{\sqrt{\pi t}(2\sqrt{t})^{2m+1}}
        \exp[-\frac{\rho^2}{4Dt}]H_{2m+1}(\frac{\rho}{\sqrt{4Dt}}),
\end{eqnarray*}
are useful.  In the above equation, $H_n(x)$ denotes the Hermite polynomial of degree $n$.  
Then, we find
\begin{eqnarray}
  g(\rho,t) &=& C(0)\frac{\rho}{\sqrt{(4\pi Dt)^3}}\exp[-\frac{\rho^2}{4Dt}] 
               + \frac{1}{4\pi\sqrt{\pi Dt}}\frac{1}{ \sinh \rho_0} \nonumber \\
        &&\times  \sum_{m=0}^\infty \frac{1}{(2m+1)!}\left(\frac{\rho_0}{2\sqrt{Dt}}\right)^{2m+1} 
          H_{2m+1}(\frac{\rho}{\sqrt{4Dt}})\exp[-\frac{\rho^2}{4Dt}].
   \label{eq.app18}
\end{eqnarray}
\section{Solution for radial symmetric diffusion equation}

From the generating function~\cite{grad80} of the Hermite polynomial, 
\begin{eqnarray*}
  {\rm e}^{-t^2}\sinh 2xt =\sum_{m=0}^{2m+1} \frac{t^{2m+1}}{(2m+1)!} H_{2m+1}(x),
\end{eqnarray*}
and Eq.(\ref{eq.app18}), we find
\begin{eqnarray}
  f(\rho,\rho_0,t)&=& \frac{1}{\sinh\rho}{\rm e}^{-Dt} g(\rho,t)
   = 4\pi D C(0)f_1(\rho,t) +f_2(\rho,\rho_0,t), \nonumber \\
  f_1(\rho,t) &=& \frac{1}{\sqrt{(4\pi Dt)^3}}{\rm e}^{-Dt} \frac{\rho}{\sinh\rho}
                \exp[-\frac{\rho^2}{4Dt}], \nonumber \\
  f_2(\rho,\rho_0,t) &=& \frac{1}{2\pi\sqrt{4\pi Dt}}{\rm e}^{-Dt}
  \frac{\sinh(\frac{\rho_0\rho}{2Dt})}{\sinh\rho_0\sinh\rho} \exp[-\frac{\rho^2+\rho_0^2}{4Dt}].
   \label{eq.app19}
\end{eqnarray}

Now, we extend the definition region of $f(\rho,\rho_0,t)$ from $\rho>\rho_0$ to $\rho>0$. Function $f_2(\rho,\rho_0,t)$ satisfies
\begin{eqnarray*}
   \int_{0}^{\infty}d\rho \int_{0}^{\pi}d\theta \int_{0}^{2\pi}d\phi
   f_2(\rho,\rho_0,t)\sinh^2\rho \,\sin\theta =1.
\end{eqnarray*}
Function $f(\rho,\rho_0,t)$ is the probability density, and satisfies the normalization condition,
\begin{eqnarray*}
   \int_{0}^{\infty}d\rho \int_{0}^{\pi}d\theta \int_{0}^{2\pi}d\phi
   f(\rho,\rho_0,t)\sinh^2\rho \,\sin\theta =1,
\end{eqnarray*}
from which constant $C(0)$ is determined as
\begin{eqnarray*}
   C(0)=0.
\end{eqnarray*}

Then, the solution of Eq.(\ref{eq.app1}), which satisfies the initial condition (\ref{eq.app2}), is given by
 \begin{eqnarray}
   f(\rho,\rho_0,t) = \frac{1}{2\pi\sqrt{4\pi Dt}}{\rm e}^{-Dt}
  \frac{\sinh(\frac{\rho_0\rho}{2Dt})}{\sinh\rho_0\sinh\rho} \exp[-\frac{\rho^2+\rho_0^2}{4Dt}].
   \label{eq.app22}
 \end{eqnarray}

\section{Summary and discussions}

In the present paper, we have investigated the radial symmetric diffusion equation (\ref{eq.app1}) in the three dimensional hyperbolic space $H_3$ with the initial condition (\ref{eq.app2}), which corresponds to the non-zero space value at the initial stage. The solution is give by Eq.(\ref{eq.app22}). 

It should be noted that Eq.(\ref{eq.app22}) reduces to Eq.(\ref{eq.int3}) in the limit of $\rho_0\rightarrow 0$. Namely,  $f(\rho,\rho_0,t)$ satisfies, 
\begin{eqnarray*}
   \lim_{\rho_0\rightarrow 0} f(\rho,\rho_0,t) = f(\rho,t).
\end{eqnarray*}

It will be necessary to consider a physical meaning of $\rho_0$ in Eq.(\ref{eq.app22}).
In high energy $AA$ collisions, some nucleons inside the nucleus would suffer inelastic collisions more than once.  In such a case, colliding nucleons have some momentum distribution in the center of mass system. Therefore, secondary particle system produced in a nucleon-nucleon collision would have non-zero velocity even at the initial stage in the center of mass system.  Such a situation would correspond to  
the initial condition (\ref{eq.app2}).  Therefore, a sort of velocity distribution of secondary particle systems at the initial stage is included in Eq.(\ref{eq.app22}).

The radial symmetric diffusion equation in the Euclidian space corresponds to the Bessel process~\cite{ito65,pitm81}. It is generalized to the Bessel process $X^{(\alpha,c)}$ in the wide sense with index $(\alpha,c)$ in Ref.~\cite{wata75}.  The process  $X^{(\alpha,c)}$ is the stochastic process in the $\alpha$-dimensional hyperbolic space $H_\alpha$. The transition density of  $X^{(\alpha,c)}$ is denoted by $p^{(\alpha,c)}(t,x,y)$ in Ref.~\cite{wata75}, and derived from a probabilistic approach different from ours. 
In fact, the solution (\ref{eq.app22}) for the diffusion equation (\ref{eq.app1}) with the initial condition (\ref{eq.app2}) satisfies,
\begin{eqnarray*}
   f(\rho,\rho_0,t)= \frac{1}{2\pi(2D)^{3/2}}p^{(3,D)}(t,\rho/\sqrt{2D},\rho_0/\sqrt{2D}).
\end{eqnarray*}
However, relation between $\rho$ and $\rho_0$ does not appear explicitly in Ref.~\cite{wata75}.

\section*{acknowledgments}
 Authors would like to thank RCNP at Osaka university, Faculty of science, Shinshu university, and Matsumoto university for financial support.

\end{document}